\documentclass[useAMS,usenatbib,usegraphicx,onecolumn]{mn2e}

\newcommand{\eqref}[1]{(\ref{#1})}

\newcommand\BV{Brunt--V\"ais\"al\"a\ }

    \def\dd{\partial}

    \def\beq{ \begin{equation} }
    \def\eeq{ \end{equation} }
    \def\spose#1{\hbox to 0pt{#1\hss}} 
    \def\ltsim{\mathrel{\spose{\lower.5ex\hbox{$\mathchar"218$}}
         \raise.4ex\hbox{$\mathchar"13C$}}}

\newcommand{\schwz}{ {\dd  \ln P\rho ^{-\gamma} \over \dd Z}}
\newcommand{\schwR} { {\dd  \ln P\rho ^{-\gamma} \over \dd R} }

\title[Rayleigh Adjustment of Narrow Barriers]
  {Rayleigh Adjustment of Narrow Barriers in Protoplanetary Discs}

\author[C.-C. Yang and K. Menou]
  {Chao-Chin Yang$^{1,2}$
   \thanks{E-mail: cyang@amnh.org (CCY); kristen@astro.columbia.edu (KM)}
   and Kristen Menou$^{3}$\\
$^{1}$Department of Astronomy, University of Illinois,
      1002 West Green Street, Urbana, IL 61801, U.S.A.\\
$^{2}$Department of Astrophysics, American Museum of Natural History,
      Central Park West at 79th Street, New York, NY 10024, U.S.A.\\
$^{3}$Department of Astronomy, Columbia University,
      550 West 120th Street, New York, NY 10027, U.S.A.}

\begin{document}

\date{Accepted 16 November 2009. Received 13 November 2009; in original form 16 June 2009}

\pagerange{\pageref{firstpage}--\pageref{lastpage}} \pubyear{2009}

\maketitle

\label{firstpage}

\begin{abstract}
Sharp density features in protoplanetary discs, for instance at the edge of a magnetically dead zone, have recently been proposed as effective barriers to slow down or even stop the problematically fast migration of planetary cores into their central star.  Density features on a radial scale approaching the disc vertical scale height might not exist, however, since they could be Rayleigh (or more generally Solberg--H\o iland) unstable.  Stability must be checked explicitly in one-dimensional viscous accretion disc models because these instabilities are artificially eliminated in the process of reducing the full set of axisymmetric equations. The disc thermodynamics, via the entropy stratification, and its vertical structure also influence stability when sharp density features are present.  We propose the concept of Rayleigh adjustment for viscous disc models: any density feature that violates Rayleigh stability (or its generalization) should be diffused radially by hydrodynamical turbulence on a dynamical time-scale, approaching marginal stability in a quasi-continuous manner.
\end{abstract}


\section{Introduction} \label{S:intro}

In current theories of planet formation, one of the most accepted and studied scenarios is the core accretion model \citep[see reviews by][and references therein]{Lissauer93,PT06}.  In this model, a Jovian planet is built by accreting gas onto a solid planetary core.  However, before the core can grow to a critical mass of $\sim$5--15~M$_\oplus$ for runaway gas accretion to occur \citep[e.g.][]{PH96,Alibert05,Rafikov06}, type~I migration due to angular momentum exchange with the natal protoplanetary gas disc can drive the core into the host star in a short time-scale \citep*[e.g.][]{GT80,Ward97,Tanaka02,MG04} compared to a typical disc lifetime of $\la$10~Myr \citep[e.g.][]{Hart98,aS06,Hillen08}.

A proposed mechanism to withstand rapid inward type~I migration is to induce a steep radial density gradient in the gas disc \citep*{MP05,MP06,MM06,MPT07,MPT09,IL08}.  In particular, since the gas inside the orbit of a planetary core exerts a positive torque on the core whereas the gas outside exerts a negative torque, a large negative radial density gradient in the gas can generate a sufficiently positive torque to halt type~I migration.  \citet{MP05,MP06} propose that the transition region in a protoplanetary disc between hydromagnetic turbulence and dead zone can locally provide such a density barrier.  On the other hand, by including the co-rotation torque exerted by the gas in the vicinity of a protoplanet, \citet{MM06} suggest that a positive torque on the protoplanet can be generated by a positive local density gradient.  \citet{IL08} further study the possibility of a local density bump around the ice line to act as a barrier, and \citet*{SLI09} complement this model by generating mock orbital configurations of extrasolar planetary systems.

However, \citet{PL84} have already pointed out that any local variation occurring on the order of a disc scale height may render the disc Rayleigh unstable.  The sharp feature may then be washed out by the resulting hydrodynamical turbulence.  Given that a large local density gradient is necessary to impact type~I migration, it is thus critical to examine the stability of such a feature.

We examine the Rayleigh stability of local density features in \S\ref{S:rsc}.  In \S\ref{S:ra}, we introduce the concept of Rayleigh adjustment as a possible solution to deal with such features in one-dimensional viscous disc models. In \S\ref{S:gen}, we discuss the need to generalize these results by accounting for the possibly important role of entropy stratification and the vertical structure.  We conclude with some general comments in \S\ref{S:impl}.

\section{Rayleigh Stability} \label{S:rsc}

Since magnetically-coupled regions in a protoplanetary disc are unstable to the magneto-rotational instability \citep{balhaw98} and unlikely to develop sharp features, our main concern is the hydrodynamical stability of a magnetically dead zone \citep{Gam96}.  The Rayleigh stability criterion is a necessary and sufficient condition for the local axisymmetric stability of an inviscid differentially rotating fluid system \citep{Chandra61}.  It states that the specific angular momentum must monotonically increase with cylindrical distance $R$ from the central rotation axis in a flow for strict stability:
\begin{equation} \label{E:rsc}
  \frac{\partial}{\partial R}(R v_\phi) > 0,
\end{equation}
where $v_\phi$ is the azimuthal velocity of the flow.  Near the mid-plane of a protoplanetary gas disc, which is supported by the central stellar gravity and the local pressure gradient, force balance in the radial direction requires that
\begin{equation} \label{E:ss}
  v_\phi^2 = v_K^2 + \frac{R}{\rho}\frac{\partial P}{\partial R},
\end{equation}
where $v_K$ is the Keplerian velocity at $R$, $\rho$ is the gas density, $P$ is the gas pressure, and any radial infall has been neglected.  Substituting equation~\eqref{E:ss} into equation~\eqref{E:rsc} gives the stability criterion
\begin{equation} \label{E:sc1}
  v_K^2 +
  \frac{1}{R}\frac{\partial}{\partial R}
             \left(\frac{R^3}{\rho}\frac{\partial P}{\partial R}\right) > 0,
\end{equation}
where we have used $v_K \propto R^{-1/2}$.

If variations in pressure, i.e.\ density or temperature for an ideal gas law, occur on a length scale of order $R$ in a disc, the second term in equation~\eqref{E:sc1} is of the order $c_s^2$, where $c_s = \sqrt{\gamma P/ \rho}$ is the sound speed and $\gamma$ is the adiabatic index (typically $7/5$ for diatomic H$_2$ in the protoplanetary context).  In a protoplanetary disc, it is usually the case that $v_K \gg c_s$ and thus the disc is Rayleigh stable irrespective of the sign of the variation.  However, if any variation occurs on a length scale of order the disc scale height, $H = R(c_s/v_K)$, the same term becomes of the order $\left(c_s R / H\right)^2 \sim v_K^2$, assuming vertical hydrostatic balance holds as usual \citep*[e.g.][]{FKR85}.  In other words, the second term in equation~\eqref{E:sc1} is of the same order as the first term.  Therefore, the disc is prone to Rayleigh instability and sensitive to the exact profile of pressure variations.

Without loss of generality (but see \S\ref{S:gen}), we adopt an ideal gas law for the equation of state and focus our stability analysis on the disc mid-plane.  We also assume for simplicity that the sound speed $c_s$ is slowly-varying with cylindrical distance $R$, as compared to the density $\rho$: $\left|(\partial c_s / \partial R) / (\partial\rho / \partial R)\right|\ll c_s / \rho$.\footnote{Our analysis could be generalized to address a steep radial temperature profile but we note that one would expect such a feature to be reduced by horizontal radiative transport (see also \S\ref{S:gen} for the role of entropy stratification).}  Then, to leading order, the stability condition (eq.~[\ref{E:sc1}]) approximately becomes
\begin{equation} \label{E:sc2}
  \left(\frac{v_K}{c_s}\right)^2 +
  \frac{R^2}{\rho}\frac{\partial^2\rho}{\partial R^2} -
  \frac{R^2}{\rho^2}\left(\frac{\partial\rho}{\partial R}\right)^2 +
  \frac{3R}{\rho}\frac{\partial\rho}{\partial R} \ga 0.
\end{equation}

For density variations on a length scale $\sim H$, the fourth term in equation~\eqref{E:sc2} is of the order $R / H \sim v_K / c_s$ and thus may be neglected.  On the other hand, both the second and the third terms are of the order $(R / H)^2 \sim \left(v_K / c_s\right)^2$.  Therefore, both the steepness and the curvature of the density profile are important to determine the stability of the disc.

Since the third term in equation~\eqref{E:sc2} is always negative, both steeply ascending and descending density profiles may make the disc Rayleigh unstable.  For example, we consider a two-fold density drop near $R = R_0$ of the form
\begin{equation} \label{E:rho1}
  \rho(R) = \rho_0\left[\frac{3}{2} - \frac{1}{2}\tanh\left(\frac{R -
                        R_0}{\epsilon}\right) \right],
\end{equation}
where $\rho_0$ is the density after the drop while $\epsilon$ controls the width of the drop.  At $R = R_0$, the first derivative of the density profile reaches maximum while the second derivative vanishes.  Substituting equation~\eqref{E:rho1} in the stability condition~\eqref{E:sc2} and evaluating it at $R = R_0$ gives $\left(v_K / c_s\right)^2 - \left(R_0 / \epsilon\right)^2 / 9 \ga 0$, where we have ignored the fourth term in equation~\eqref{E:sc2}.  Since $H / R = c_s / v_K$, it requires that $\epsilon \ga H / 3$ and the maximum slope allowed in this case is $|\partial\rho / \partial R| \sim 3\rho_0 / 2H$.  This illustrates how Rayleigh stability sets an upper limit on the steepness of a density profile.

Furthermore, the second term in equation~\eqref{E:sc2} indicates that a disc may be Rayleigh unstable where the second derivative of the density profile is negative.  In regions where the profile is concave down, the term sets a lower limit to the radius of curvature.  In particular, a cusp pointing upward cannot be stable due to an infinite negative second derivative.  Smoothness is a necessary condition in these regions.  Locations where pressure or density maxima occur in a protoplanetary disc are of particular interest in that solid grains are captured, but Rayleigh instability may also come into question at these locations in a nontrivial manner.  To illustrate this point, we consider a density bump with a Gaussian form,
\begin{equation} \label{E:rho2}
  \rho(R) = \rho_0 \left\{1 + \frac{H}{\epsilon}
            \exp\left[-\frac{\left(R - R_0\right)^2}
            {\epsilon^2}\right]\right\}.
\end{equation}
Superposed on a disc of uniform mid-plane density $\rho_0$, the bump encloses a constant mass (on the order of the mass in an annulus of mid-plane density $\rho_0$ and width $H$ at $R = R_0$) and has a width proportional to $\epsilon$, provided that $\epsilon \ll R_0$.  The most negative second derivative occurs at the peak, where the first derivative vanishes.  Evaluating the stability condition~\eqref{E:sc2} at $R = R_0$ results in $\left(v_K / c_s\right)^2 - 2H\left(R_0 / \epsilon\right)^2 / (H + \epsilon) \ga 0$.  Since $H / R = c_s / v_K$, the minimum width allowed for stability is $\epsilon \sim H$.  Therefore, a local density maximum cannot be Rayleigh stable if the width of the bump is less than about a disc scale height.

\section{Rayleigh Adjustment} \label{S:ra}

Once a certain region in a protoplanetary disc reaches a state of Rayleigh instability, we conjecture that the ensuing hydrodynamical turbulence will redistribute material in the disc in such a way that any narrow feature will be smoothed and broadened till the disc recovers a state that is (marginally) Rayleigh stable.  The time-scale $t_R$ for a disc to recover a Rayleigh-stable state might be close to the dynamical time-scale of the disc, i.e.\ $t_R \ga \Omega_{\rm K}^{-1}$, where $\Omega_{\rm K}$ is the Keplerian angular velocity, although a slower growth rate may be expected as one approaches marginal stability.  By contrast, standard $\alpha$-model of thin accretion discs has a viscous time-scale $t_\nu \sim R^2 / \alpha H c_s$ \citep[e.g.][]{FKR85}.  Given that $\alpha < 1$ and $H \ll R$, $t_\nu > (R / H)^2 t_R \gg t_R$.  In other words, any model of viscously-evolving protoplanetary discs should maintain Rayleigh stability quasi-steadily.

If the condition $t_\nu \gg t_R$ holds, it is useful to implement an adjustment scheme to guarantee that a disc model remains Rayleigh quasi-stable at all time.  In evolutionary models like those of \citet{MPT07,MPT09} or \citet{IL08}, for instance, this kind of adjustment could be applied at any time-step whenever Rayleigh instability is detected, before one proceeds with the migration torque calculation.  The purpose of the adjustment is to capture the final state of the adjustment process when the disc just restores its Rayleigh stability and rotational equilibrium (implied by eq.~[\ref{E:ss}]), without any detailed knowledge of the Rayleigh hydrodynamical turbulence occurring on small temporal or spatial scales.

In the hydrodynamical context of interest here, Rayleigh adjustment must involve some level of mass and angular momentum redistribution in the disc.  To the extent that the disc reaches a new state of rotational equilibrium (described by eq.~[\ref{E:ss}]) after Rayleigh adjustment, however, the adjustment itself can be described as a transition from one state of rotational equilibrium to another such state, under the action of mass redistribution.  Therefore, this process may be phenomenologically described by the radial mass diffusion equation
\begin{equation} \label{E:diff}
  \frac{\partial\Sigma}{\partial t} =
  \frac{1}{R}\frac{\partial}{\partial R}
             \left(R D\frac{\partial\Sigma}{\partial R}\right),
\end{equation}
where $\Sigma$ is the vertically-integrated column density and $D$ is the mass diffusion coefficient.  We use $\Sigma$ instead of $\rho$ under the hypothesis that the transport is vertically global.  Whether or not the disc evolution due to Rayleigh turbulence can be described by fast radial mass diffusion remains to be demonstrated by full hydrodynamical simulations.  In this regard, equation~\eqref{E:diff} can only be deemed to be an approximation to the actual adjustment process.

We hereby describe a simple algorithm to implement Rayleigh adjustment, using the diffusion equation~\eqref{E:diff}.  As a first step, we ignore any spatial variations and assume that $D$ is a constant to focus our discussion on how to include Rayleigh adjustment into viscously-evolving disc models.  The equation can be discretized as
\begin{equation} \label{E:disc}
  \Sigma_j^{n+1} =
  \Sigma_j^n +
  \frac{D\Delta t}{\Delta R^2}
  \left[\left(\Sigma_{j+1}^n - 2\Sigma_j^n + \Sigma_{j-1}^n\right) +
        \frac{\Delta R}{2R_j}
        \left(\Sigma_{j+1}^n - \Sigma_{j-1}^n\right)\right],
\end{equation}
where $\Sigma_j^n$ is the density at $R = R_j$ and step $n$, $\Delta t$ is the time increment of the adjustment scheme, and $\Delta R = R_{j+1} - R_j$ is the grid spacing.  We have adopted a regular mesh in equation (8) for simplicity.  From von Neumann stability analysis, $\Delta t \le \Delta R^2 / 2D$.  Substituting $\Delta t = \Delta R^2 / 2D$ in equation~\eqref{E:disc} gives
\begin{equation} \label{E:relax}
  \Sigma_j^{n+1} = \frac{1}{2}\left(\Sigma_{j+1}^n +
                   \Sigma_{j-1}^n\right) + \frac{\Delta
                   R}{4R_j}\left(\Sigma_{j+1}^n -
                   \Sigma_{j-1}^n\right).
\end{equation}
A Rayleigh unstable density profile can then be treated as an initial guess and equation~\eqref{E:relax} can be iterated to the point when the density first becomes everywhere Rayleigh stable, under the constraint that rotational equilibrium is satisfied throughout (eq.~[\ref{E:ss}]).  This corresponds to the completion of the turbulent redistribution process induced by Rayleigh instability.  Note that the magnitude of the diffusion coefficient $D$ plays no role in determining the final state of the Rayleigh adjustment, which can be seen in equation~\eqref{E:relax}, since we have chosen the maximum time-step possible to minimize the number of iterations needed to find the adjusted state.  We have verified that the final profile does not depend on the time-step, as long as the von Neumann stability condition is satisfied, and that the total mass of the disc is conserved.

We emphasize that since the diffusion time-scale of this process is about $L^2 /D$, where $L$ is the characteristic length scale of the density profile, sharp Rayleigh-unstable density features (with $L \ll R$) will diffuse much more rapidly than any density structure on large scales.  Therefore, in first approximation, the global disc structure will remain largely unaffected by this Rayleigh adjustment even for a globally constant $D$.  Note that such a quasi-instantaneous adjustment scheme can only be justified if the process leading to Rayleigh instability (e.g.\ viscous mass pile-up) operates much more slowly than the adjustment itself.

\begin{figure}
\begin{center}
\includegraphics[width=0.8\textwidth]{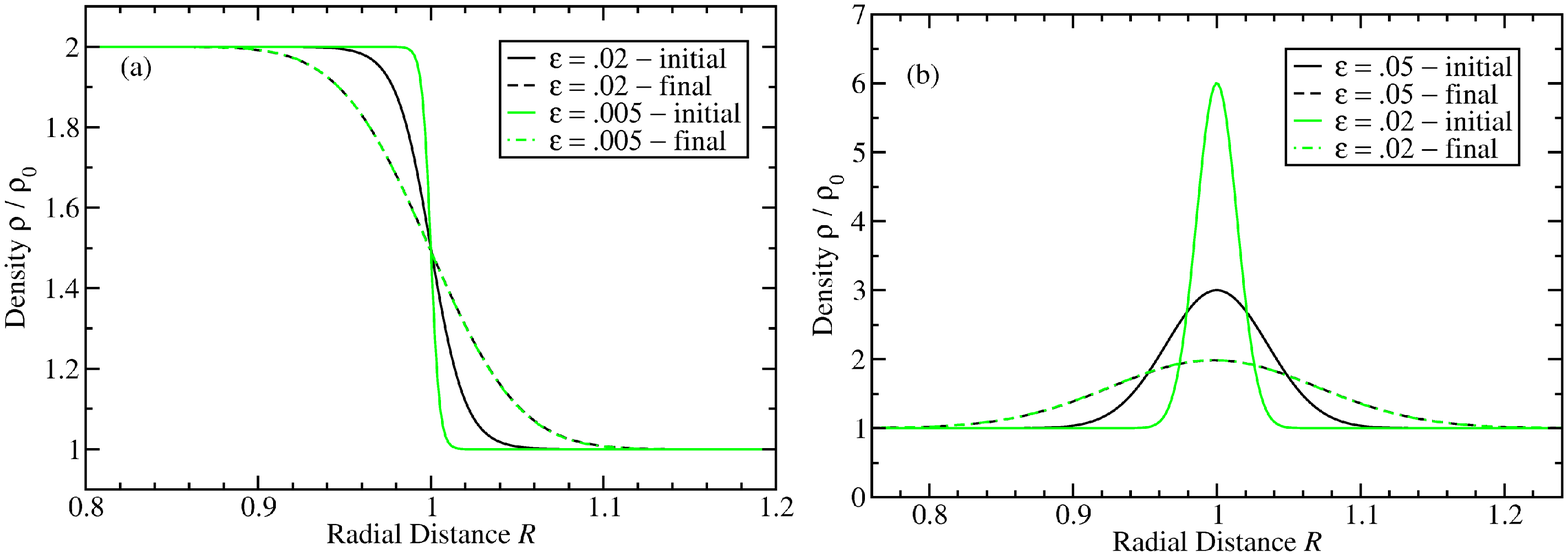}
\caption{Initially Rayleigh unstable (\emph{solid} lines) and final Rayleigh-stable (\emph{dashed} and \emph{dot-dashed} lines) profiles for (a)~a density drop (eq.~[\ref{E:rho1}]) and (b)~a density bump (eq.~[\ref{E:rho2}]) processed through the Rayleigh adjustment scheme discussed in \S\ref{S:ra}.  With the constant diffusion coefficient adopted in this algorithm, the final relaxed density profiles do not depend on the width of the initially unstable profiles.}
\label{Fi:sample}
\end{center}
\end{figure}

As a demonstration, we have implemented this adjustment scheme on the two model density profiles, equations~\eqref{E:rho1} and~\eqref{E:rho2}, previously discussed in \S\ref{S:rsc}.  For concreteness, we adopt $R_0 = 1$ and $H = 0.1$.  The parameter $\epsilon$ is chosen such that the density profile is Rayleigh unstable.  The profile is discretized with 1000 grid points in the range $0.5 \le R \le 2$ and diffused via equation~\eqref{E:relax}.  As shown in Fig.~\ref{Fi:sample}, profiles of different initial widths relax to the same enlarged width at marginal Rayleigh stability, which is roughly one disc scale height, $H$.  Even though these results are based on a constant mass diffusion coefficient, they are consistent with the expectation that the width of any density profile is intrinsically limited to $\sim H$ by Rayleigh stability.  One expects any variation narrower than $\sim H$ to be smoothed and broadened, although details of this process may depend on the specific Rayleigh adjustment prescription adopted.  We note that, in addition to redistributing mass and angular momentum, localized hydrodynamical turbulence in the disc could also directly influence the migration process \citep*[e.g.][]{Laugh04,Nelson05,JGM06,Oishi07}, an effect which is not captured by our adjustment scheme.  Potential consequences for the disc thermodynamics have also been ignored.

We note that the mass diffusion equation~\eqref{E:diff} employed here conserves mass in the disc but it does not conserve angular momentum during the adjustment process, under the assumption that rotational equilibrium holds before and after Rayleigh adjustment, as specified by equation~\eqref{E:ss}.  At first, this would seem to be a serious limitation of the simple mass redistribution scheme proposed here.  One may be tempted to improve such a scheme by using instead a more standard viscous equation for mass and angular momentum transport in a thin disc, of the form
\begin{equation} \label{E:dead}
  \frac{\partial\Sigma}{\partial t} = \frac{3}{R} \frac{\partial
  }{\partial R}\left[ R^{1/2} \frac{\partial }{\partial R} (\nu \Sigma
  R^{1/2}) \right],
\end{equation}
where $\nu$ is the effective disc viscosity \citep[e.g.][]{FKR85}.  However, even though this equation does conserve mass and angular momentum in a Keplerian disc, it would be as inadequate as equation~\eqref{E:diff} if used in a Rayleigh adjustment scheme.  Indeed, equation~\eqref{E:dead}, or its generalizations for non-Keplerian rotation laws, assumes that the disc angular velocity profile is fixed in time.  It is precisely the breakdown of this assumption during Rayleigh adjustment, $\partial v_\phi / \partial t \neq 0$, going from one state of rotational equilibrium to another such equilibrium, which prevents this entire class of equations from accurately tracking angular momentum conservation during a Rayleigh adjustment episode.  To the extent that one is interested in describing the global evolution of discs in rotational equilibrium with approximate equations such as equation~\eqref{E:dead}, one may then choose to consider the small violations in angular momentum conservation occurring during Rayleigh adjustments as an acceptable source of errors, given that it is an unspecified detail at the level of approximation of the viscous theory.  This, in our view, motivates the use of the simplest possible mass diffusion scheme, under the constraint of rotational equilibrium, as we have proposed here with equation~\eqref{E:diff}.

To summarize, the general outline of a Rayleigh adjustment procedure would be as follows. At each time-step of a one-dimensional evolutionary disc model, evaluate the stability of the disc profile using equation~\eqref{E:sc1} or a generalization (see \S\ref{S:gen} below).  If the profile is stable, move on to the next time-step of the evolution.  Otherwise, use the unstable profile as the initial condition and iterate equation~\eqref{E:relax}, or a more sophisticated form of equation~\eqref{E:diff}, till the profile has relaxed to marginal stability.  Then use the relaxed profile to resume the evolutionary model.  Two important assumptions underlying this scheme are (i)~that the agent driving the disc viscous evolution acts on a time-scale much longer than the disc dynamical time, (ii)~that the relaxed density profile is independent of the pre-adjusted profile, as illustrated in Fig.~\ref{Fi:sample} for the simple scheme proposed here.

\section{Role of Entropy Stratification and Vertical Structure} \label{S:gen}

So far, our discussion has focused exclusively on the Rayleigh stability of a purely radial stratification of angular momentum in the disc mid-plane.  More generally, entropy stratification will also be present in the protoplanetary disc and contribute to its axisymmetric stability if sharp density features exist.  The full generalization of Rayleigh's stability analysis in the presence of entropy stratification leads to the two necessary and sufficient Solberg--H\o iland criteria for axisymmetric stability \citep[e.g.][]{Tassoulbook}:
\beq\label{hoilad1}
N_R^2 + N_Z^2
+ {1\over R^3} {\dd R^4\Omega^2\over \dd R} > 0,
\eeq
\beq\label{hoilad2}
\left( - {\dd P\over \dd Z} \right) \, \left( {1 \over R^3}
{\dd R^4 \Omega^2\over\dd R} \schwz - {1 \over R^3} 
{\dd R^4\Omega^2\over\dd Z}\schwR \right) > 0,
\eeq
where 
\beq\label{hoilad3} N_R^2= -{1 \over \gamma\rho} {\dd P \over \dd R} {\dd \ln
P\rho ^{-\gamma} \over \dd R}, \qquad N_Z^2= -{1 \over \gamma\rho}
{\dd P \over \dd Z} {\dd \ln P\rho ^{-\gamma} \over \dd Z}, 
\eeq 
$\Omega = v_\phi / R$ is the angular velocity, and $Z$ is the vertical coordinate along the rotation axis.  The term $N_R^2 + N_Z^2$ is the sum of the cylindrical radial and vertical components of the squared \BV frequency.

Equations~\eqref{hoilad1}--\eqref{hoilad2} make it clear that the stratifications of entropy and angular momentum both contribute to axisymmetric stability.  Since these criteria were derived with respect to any linear axisymmetric perturbation in the $(R,Z)$ plane, they address the cylindrical radial and the vertical components of both stratifications.  In the presence of a sharp radial density feature, the radial component of the entropy stratification, which is normally negligible in a disc, can make a significant contribution to local stability.  To illustrate this possibility, let us consider a case again where the sound speed $c_s$ is slowly-varying with cylindrical distance $R$ as compared to the density, and let us restrict perturbations to be purely cylindrical radial so that vertical stratification plays no role ($\dd / \dd Z \to 0$ in eqs.~[\ref{hoilad1}]--[\ref{hoilad3}]).  In this limit, axisymmetric stability reduces to the much simpler criterion
\begin{equation} 
N_R^2 + {1\over R^3} {\dd R^4\Omega^2\over \dd R} > 0,
\end{equation}
which is the equivalent of equation~(\ref{E:rsc}) with an additional contribution from radial entropy stratification.  According to the same scaling analysis as used in \S\ref{S:rsc} for equation~(\ref{E:sc2}), in the presence of a radial density gradient with length scale $H$, neglecting any radial temperature gradient on such scales,
\begin{equation} 
N_R^2 \simeq \frac{c_s^2}{H^2} \left( 1-\frac{1}{\gamma} \right)
\simeq \frac{v_K^2}{R^2} \left( 1-\frac{1}{\gamma} \right).
\end{equation}
The magnitude of the radial entropy stratification, as measured by $N_R^2$, could thus be comparable to that of the angular momentum stratification and contribute to the radial stability of the disc configuration.

This suggests that, in the presence of sharp density gradients, it may no longer be possible to ignore the disc thermodynamics for stability considerations.  Perhaps even more importantly, equations~\eqref{hoilad1}--\eqref{hoilad2} show that the disc radial stability can no longer be studied independently of its vertical structure.  In the presence of radial density features with length scale $\sim H$, radial and vertical gradients become comparable in magnitude.  The disc stability is then determined by a detailed balance between the radial and vertical stratifications of angular momentum and entropy (eqs.~[\ref{hoilad1}]--[\ref{hoilad2}]).  This makes the stability analysis significantly more complicated than considered so far, especially in the case of dead zones with a poorly known vertical structure.  It may also imply that a more elaborate treatment than the simple Rayleigh adjustment scheme proposed in \S\ref{S:ra} is required for unstable discs.

\section{Discussion and Conclusion} \label{S:impl}

The possibility that discs with sharp radial density features may be subject to Rayleigh instability has already been noted by various authors \citep[e.g.][]{PL84,MPT07,MPT09}.  We have reconsidered this issue here and have suggested that stability be checked explicitly in viscous evolutionary disc models which develop such features.  In viscous disc models, the full set of axisymmetric equations is reduced to a single radial diffusion equation of the type shown in equation~\eqref{E:dead}.  Typically, the equation reduction is achieved by imposing a steady angular velocity profile in the disc, $\Omega(R,t) = \Omega(R)$ (often chosen to be Keplerian, like in eq.~[\ref{E:dead}]), and by neglecting the radial acceleration term in the radial momentum equation.  These are simplifications which contribute to eliminating the class of axisymmetric instabilities discussed here.  If sharp density features are to be considered as viable solutions to slow down or even stop protoplanetary migration, with many important consequences for planet formation scenarios \citep[e.g.][]{MPT07,MPT09,IL08,SLI09}, a careful examination of stability in viscously evolved models seems warranted.

The various arguments we have put forward are not meant to imply that the sharp density features considered in various works \citep[e.g.][]{MPT07,MPT09,IL08,SLI09} are axisymmetrically unstable.  However, they suggest that an approximate scaling analysis based on a representative length-scale $\sim H$ may not be sufficiently accurate to evaluate the axisymmetric stability of a disc, which depends on the detailed shape of the density profile.  In addition, since both the vertical structure and the entropy stratification of the disc can influence its stability when sharp features are present, a careful treatment of the disc thermodynamics may be required.  The extent to which density features can or cannot grow to be as sharp as a disc scale height may significantly affect planetary migration because differential Lindblad torques are also applied over a length-scale $\sim H$ \citep[e.g.][]{Ward97,MPT07} and co-rotation torques are very sensitive to the local density \citep{MM06} and thermodynamic \citep{PaaPap08} conditions.  For those discs which develop unstable density profiles as a result of slow viscous evolution, for instance mass pile-up in a magnetically dead zone, we have proposed that Rayleigh adjustment schemes be implemented.  Although more detailed work would be needed to incorporate such a scheme within the framework of a viscous disc solver, our goal is to suggest that this type of adjustment schemes can be used to evolve more consistently discs with sharp features on viscous evolutionary time-scales.

Finally, we remark that our discussion has been restricted to the stability of axisymmetric perturbations in protoplanetary discs.  It has been suggested that discs with narrow density features are also susceptible to non-axisymmetric instabilities \citep{LF00,LC01,LJ09}.  To the extent that such instabilities lead to mass diffusion on a fast dynamical time-scale, it should be possible to model their effects on disc evolutionary time-scales with an adjustment scheme similar to the one proposed here.

\section*{Acknowledgments}
We thank an anonymous reviewer for comments which led us to clarify the content of this manuscript.  This work was supported by NASA OSS grant \#NNX07AI74G.


\bsp

\label{lastpage}


\begin{thebibliography}{99}

\bibitem[\protect\citeauthoryear{Alibert et al.}{2005}]{Alibert05}
Alibert Y., Mordasini C., Benz W., Winisdoerffer C., 2005, A\&A, 434, 343

\bibitem[\protect\citeauthoryear{Balbus \& Hawley}{1998}]{balhaw98}
Balbus S.~A., Hawley J.~F., 1998, RvMP, 70, 1

\bibitem[\protect\citeauthoryear{Chandrasekhar}{1961}]{Chandra61}
Chandrasekhar S., 1961, Hydrodynamic and Hydromagnetic Stability.\
Clarendon Press, Oxford

\bibitem[\protect\citeauthoryear{Frank, King \& Raine}{Frank et
al.}{2002}]{FKR85} Frank J., King A.~R., Raine D.~J., 2002, Accretion
Power in Astrophysics. Third Edition\ Cambridge Univ. Press, New York,
NY

\bibitem[\protect\citeauthoryear{Gammie}{1996}]{Gam96}
Gammie C.~F., 1996, ApJ, 457, 355

\bibitem[\protect\citeauthoryear{Goldreich \& Tremaine}{1980}]{GT80}
Goldreich P., Tremaine S., 1980, ApJ, 241, 425 

\bibitem[\protect\citeauthoryear{Hartmann et al.}{1998}]{Hart98}
Hartmann L., Calvet N., Gullbring E., D'Alessio P., 1998, ApJ, 495, 385 

\bibitem[\protect\citeauthoryear{Hillenbrand}{2008}]{Hillen08}
Hillenbrand L.~A., 2008, PhST, 130, 014024

\bibitem[\protect\citeauthoryear{Ida \& Lin}{2008}]{IL08}
Ida S., Lin D.~N.~C., 2008, ApJ, 685, 584

\bibitem[\protect\citeauthoryear{Johnson, Goodman \& Menou}{Johnson et al.}{2006}]{JGM06}
Johnson E.~T., Goodman J., Menou K., 2006, ApJ, 647, 1413 

\bibitem[\protect\citeauthoryear{Laughlin, Steinacker \& Adams}{Laughlin et al.}{2004}]{Laugh04}
Laughlin G., Steinacker A., Adams F.~C., 2004, ApJ, 608, 489 

\bibitem[\protect\citeauthoryear{Li et al.}{2000}]{LF00}
Li H., Finn J.~M., Lovelace R.~V.~E., Colgate S.~A., 2000, ApJ, 533, 1023 

\bibitem[\protect\citeauthoryear{Li et al.}{2001}]{LC01}
Li H., Colgate S.~A., Wendroff B., Liska R., 2001, ApJ, 551, 874 

\bibitem[\protect\citeauthoryear{Lissauer}{1993}]{Lissauer93}
Lissauer J.~J., 1993, ARA\&A, 31, 129 

\bibitem[\protect\citeauthoryear{Lyra et al.}{2009}]{LJ09}
Lyra W., Johansen A., Zsom A., Klahr H., Piskunov N., 2009, A\&A, 497, 869 

\bibitem[\protect\citeauthoryear{Masset et al.}{2006}]{MM06} 
Masset F.~S., Morbidelli A., Crida A., Ferreira J., 2006, ApJ, 642, 478 

\bibitem[\protect\citeauthoryear{Matsumura \& Pudritz}{2005}]{MP05}
Matsumura S., Pudritz R.~E., 2005, ApJ, 618, L137

\bibitem[\protect\citeauthoryear{Matsumura \& Pudritz}{2006}]{MP06}
Matsumura S., Pudritz R.~E., 2006, MNRAS, 365, 572

\bibitem[\protect\citeauthoryear{Matsumura, Pudritz \& Thommes}{Matsumura et al.}{2007}]{MPT07}
Matsumura S., Pudritz R.~E., Thommes E.~W., 2007, ApJ, 660, 1609 

\bibitem[\protect\citeauthoryear{Matsumura, Pudritz \& Thommes}{Matsumura et al.}{2009}]{MPT09}
Matsumura S., Pudritz R.~E., Thommes E.~W., 2009, ApJ, 691, 1764 

\bibitem[\protect\citeauthoryear{Menou \& Goodman}{2004}]{MG04}
Menou K., Goodman J., 2004, ApJ, 606, 520 

\bibitem[\protect\citeauthoryear{Nelson}{2005}]{Nelson05}
Nelson R.~P., 2005, A\&A, 443, 1067 

\bibitem[\protect\citeauthoryear{Oishi, Mac Low \& Menou}{Oishi et al.}{2007}]{Oishi07}
Oishi J.~S., Mac Low M.-M., Menou K., 2007, ApJ, 670, 805 

\bibitem[\protect\citeauthoryear{Paardekooper \& Papaloizou}{2008}]{PaaPap08}
Paardekooper S.-J., Papaloizou J.~C.~B., 2008, A\&A, 485, 877 

\bibitem[\protect\citeauthoryear{Papaloizou \& Lin}{1984}]{PL84}
Papaloizou J., Lin D.~N.~C., 1984, ApJ, 285, 818 

\bibitem[\protect\citeauthoryear{Papaloizou \& Terquem}{2006}]{PT06}
Papaloizou J.~C.~B., Terquem C., 2006, RPPh, 69, 119 

\bibitem[\protect\citeauthoryear{Pollack et al.}{1996}]{PH96}
Pollack J.~B., Hubickyj O., Bodenheimer P., Lissauer J.~J., Podolak M.,
Greenzweig Y., 1996, Icar, 124, 62 

\bibitem[\protect\citeauthoryear{Rafikov}{2006}]{Rafikov06}
Rafikov R.~R., 2006, ApJ, 648, 666 

\bibitem[\protect\citeauthoryear{Schlaufman, Lin \& Ida}{Schlaufman et al.}{2009}]{SLI09}
Schlaufman K.~C., Lin D.~N.~C., Ida S., 2009, ApJ, 691, 1322 

\bibitem[\protect\citeauthoryear{Sicilia-Aguilar et al.}{2006}]{aS06}
Sicilia-Aguilar A., Hartmann L., Calvet, N.\ et al., 2006, ApJ, 638, 897 

\bibitem[\protect\citeauthoryear{Tanaka, Takeuchi \& Ward}{Tanaka et al.}{2002}]{Tanaka02}
Tanaka H., Takeuchi T., Ward W.~R., 2002, ApJ, 565, 1257 

\bibitem[\protect\citeauthoryear{Tassoul}{1978}]{Tassoulbook}
Tassoul J.~L., 1978, Theory of Rotating Stars.\
Princeton Univ. Press, Princeton, NJ

\bibitem[\protect\citeauthoryear{Ward}{1997}]{Ward97}
Ward W.~R., 1997, Icar, 126, 261 

\end{thebibliography}
\end{document}